\documentclass[10pt,twocolumn,aps,prd,preprintnumbers,showpacs,superscriptaddress,nofootinbib,amsmath,amssymb,floats,floatfix, nofootinbib,notitlepage]{revtex4-1}

\usepackage{graphicx}
\usepackage{hyperref}
\usepackage{subfigure}
\usepackage{multirow}
\usepackage[toc,page]{appendix}
\usepackage[normalem]{ulem}
\usepackage{adjustbox}
\usepackage{latexsym}
\usepackage{amsmath}
\usepackage{amssymb}
\usepackage{amsfonts}
\usepackage{times}
\usepackage{dcolumn}
\usepackage{bm}
\usepackage{tikz}
\usepackage{bigints}
\usepackage{array,tabularx,multirow,booktabs}
\usepackage{tabularx}
\usepackage{multirow}
\usepackage[tracking=true]{microtype}
\SetTracking{}{500}
\SetTracking{encoding={*}, shape=sc}{40}

\begin{document}
\title{Weak deflection angle of a dirty black hole}
\author{Reggie C. Pantig}
\email{reggie.pantig@dlsu.edu.ph}
\affiliation{Physics Department, De La Salle University-Manila, 2401 Taft Ave., 1004 Manila Philippines}
\author{Emmanuel T. Rodulfo}
\email{emmanuel.rodulfo@dlsu.edu.ph}

\begin{abstract}
In this paper, we present the weak deflection angle in a Schwarzschild black hole of mass $m$ surrounded by the dark matter of mass $M$ and thickness $\Delta r_{s}$. The Gauss-Bonnet theorem, formulated for asymptotic spacetimes, is found to be ill-behaved in the third-order of $1/\Delta r_{s}$ for very large $\Delta r_{s}$. Using the finite-distance for the radial locations of the source and the receiver, we derived the expression for the weak deflection angle up to the third-order of $1/\Delta r_{s}$ using Ishihara (\textit{et al.}) method. The result showed that the required dark matter thickness is $\sim2\sqrt{3mM}$ for the deviations in the weak deflection angle to occur. Such thickness requirement is better by a factor of 2 as compared to the deviations in the shadow radius ($\sim\sqrt{3mM}$). It implies that the use of the weak deflection angle in detecting dark matter effects in one's galaxy is better than using any deviations in the shadow radius.
\end{abstract}

\maketitle

\section{Introduction} \label{sec1}
One of the fruitful achievements of the human mind is the general theory of relativity by Albert Einstein, which relates the phenomena of gravitation to the geometry of spacetime. One consequence of the theory is the existence of black holes that have long remained to be a theoretical construct, until the release of the first image of the black hole shadow at the center of the M87 galaxy on April 10, 2019 \cite{Akiyama2019a}.

At present, dark matter is another entity that remains so mysterious and elusive. The $\Lambda$CDM model of cosmology suggests that the content of our universe is made up of 27$\%$ dark matter, which constitutes 85$\%$ of the total mass \cite{Jarosik2011}. Using Earth-based laboratories, several scientists attempted to detect dark matter by direct means, but they gave inconsistent results with each other. (see \cite{Bernabei2010} and \cite{Aprile2010}). Even long before these experiments, indirect search through dark matter annihilation also revealed null results \cite{Desai2004, Peirani2004}. It proves that dark matter detection is more difficult compared to gravitational wave detection \cite{Abbott2016}.

Emerging theoretical efforts have shown alternative means of possible dark matter detection using the changes in the silhouette of a black hole. The spacetime of pure dark matter and black hole were combined rigorously by Xu in Ref. \cite{xu2018black}, and this formalism of using dark matter density profiles to black hole spacetime is applied immediately to our own and M87 galaxy \cite{hou2018black, Jusufi2019}. 
Several studies are also present that analyze the black hole shadow and deflection angle under the influence of dark matter modeled as a perfect fluid \cite{hou2018rotating,haroon2019shadow}. However, Konoplya in Ref. \cite{Konoplya2019} considered these black hole metrics as highly model-dependent. Using a less model-dependent and agnostic view on dark matter, he estimated the condition for dark matter to have some notable effect on the shadow radius.

Not only the study of shadow from various black hole models has gained much attention from many researchers, but also the study of gravitational lensing (GL). GL has proved useful in probing many astrophysical objects. Long ago, it is used to probe the coevolution of supermassive black holes (SMBH) and galaxies \cite{Peng2006}. It is also used to probe exotic entities like dark matter \cite{Trimble1987,Metcalf2001,Metcalf2002} that permeates a whole galaxy, and even galaxy clusters \cite{Hoekstra2008,Ellis2010}. Astrophysical black holes, which can also be approximately described by the Schwarzschild metric for useful estimates, 
are studied extensively in terms of lensing and the relativistic images that it produces \cite{Virbhadra:1999nm, Virbhadra:2008ws}.

Perhaps the most popular way to obtain the weak deflection angle is by using the Gauss-Bonnet theorem (GBT) introduced in Ref. \cite{Gibbons2008}. Since then, the GBT has been proven very useful in calculating the weak deflection angle of various black hole models that demonstrate asymptotic behavior \cite{Ovgun2018,Ovgun2018a,Ovgun2018b,Ovgun2019a,Ovgun2019c,Jusufi2018,Javed}. Other notable studies also existed that includes dark matter, phantom matter, and quintessential energy, \cite{Ovgun2019, Ovgun2019b, haroon2019shadow, DeLeon2019} on their analysis of the weak deflection angle. Gravitational lensing by exotic matter or energy, which possibly deviate from some well-known equation of state, is also explored in \cite{Kitamura2014, Nakajima2014, Izumi2013}.

Calculation of the weak deflection angle for non-asymptotic spacetimes seems problematic using the GBT as far as its asymptotic form is concerned. Recently, an improved analysis in Ref. \cite{Ishihara2016} made the calculation possible by considering the finite distances of the source and receiver. However, the correspondence between the GBT and finite-distance involves the brute force of evaluating integrals, which contain the orbit equation. This method, whether the spacetime is asymptotic or not, axially-symmetric or not \cite{Ono2019, Li2020a, Zhang2019a}, is also useful in strong gravitational lensing regime \cite{Ishihara2017, Azreg-Ainou2017}.

It is interesting to investigate the effect of dark matter configuration used in Ref. \cite{Konoplya2019} on a different black hole phenomenon: weak gravitational lensing. In particular, this paper seeks to derive the expression for the weak deflection angle caused by the main property of dark matter - its mass. Although that the deviation in the shadow radius already gives us the idea that null geodesics are affected, it is interesting to find out whether if the deviation in the black hole's weak deflection angle offers a better condition for dark matter detection.

We organize the paper as follows: Sect. \ref{sec2} introduces the dirty Schwarzschild metric alongside with the description of the simple dark matter model in consideration. In Sect. \ref{sec3}, we present three possible cases that show the different positions of the source and receiver relative to the dark matter distribution and utilize the GBT. In Sect. \ref{sec4}, we proceed to calculate the weak deflection angle by assuming some finite-distance of the source and the receiver using the Ishihara (\textit{et al.}) method. Lastly, in Sect. \ref{sec5}, we summarize the results and indicate some possible future research direction. The metric signature in this study is +2, and we use $G=c=1$.

\section{Dirty Schwarzschild black hole} \label{sec2}
The term "dirty" black hole has its roots in Ref. \cite{Visser1992,Visser1993,Macedo2016,Leung2018,Krauss1996} which describes a black hole surrounded by some astrophysical environment. These dirty black holes can be categorized as follows: (a) those that came from a specific Lagrangian, or a certain field theory which results to a metric as the Einstein field equation is solved (for examples, see Refs. \cite{Shapere1991, Dowker1992, Gibbons1988, Allen1990, Galtsov1989, Lahiri1992}), (b) generic black hole metrics with sufficient generality \cite{Nielsen2019} that came from some hypothetical configuration (or derived from empirical data) of astrophysical environment, and (c) dirty black holes which came from solutions to a non-Einsteinian theory such as pseudo-complex general relativity \cite{Moffat1979, Hess2009, Mann1982}. 

Here, we give a brief overview and describe the dirty Schwarzschild black hole used in Ref. \cite{Konoplya2019}, where the author studied dark matter effects on the photonsphere and black hole shadow. The astrophysical environment in consideration is a spherical shell of dark matter described only by its mass M, inner radius $r_{s}$, and thickness $\Delta r_{s}$ (the subscript $s$ denotes shell). Further, the dark matter mass $M$ is treated as an additional effective mass to the black hole while maintaining its non-interaction with the electromagnetic field. Since the physical parameters of the dark matter shell are hypothetical, such a dirty black hole falls to the second category described above. The generic metric produced still has sufficient generality because the black hole itself came from the vacuum solution of the Einstein equation.

\begin{figure}
    \centering
    \includegraphics[width=\linewidth]{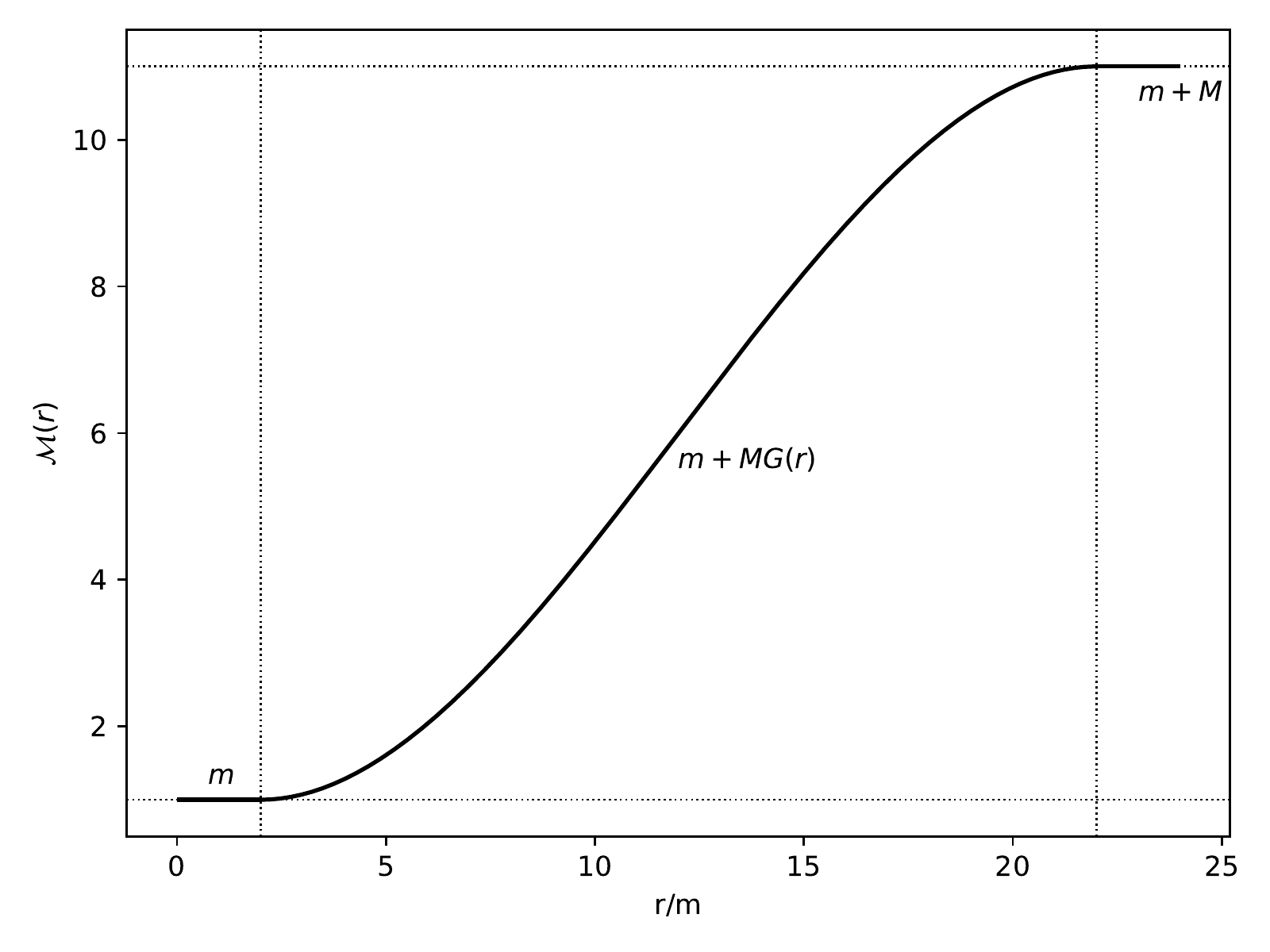}
    \caption{Example of a choice for mass function. Here, $M=10m$, $r_{s}=2m$, $\Delta r_{s}=20m$, $m=1$.}
    \label{fig1}
\end{figure}

One can then assume a piecewise function to impose three domains \cite{Konoplya2019}:
\begin{equation} \label{massfunc}
\mathcal{M}(r)=\begin{cases}
m, & r<r_{s};\\
m+ M G(r), & r_{s}\leq r\leq r_{s} +\Delta r_{s};\\
m+ M, & r>r_{s}+\Delta r_{s}
\end{cases}
\end{equation}
where
\begin{equation} \label{e2}
G(r)=\left(3-2\frac{r-r_{s}}{\Delta r_{s}}\right)\left(\frac{r-r_{s}}{\Delta r_{s}}\right)^{2}.
\end{equation}
The expression for $G(r)$ is chosen so that $\mathcal{M}(r)$ and $\mathcal{M}'(r)$ are continuous (see Fig. \ref{fig1}). Note also the possibility of $M<0$.

The Schwarzschild metric is one of the famous, yet simplest, vacuum solution of the Einstein field equations. Consider surrounding it with a spherical shell of dark matter whose parameters are described by Eq. \eqref{e2}. Then we have
\begin{equation} \label{e3}
d{s}^2 = -f(r)dt^2 +f(r)^{-1}dr^2 +r^2\left(d\theta^2+\sin^2 \theta d\phi^2\right)
\end{equation}
where the metric function $f(r)$ is now given by
\begin{equation} \label{e4}
f(r)=1-\frac{2\mathcal{M}(r)}{r}.
\end{equation}

\begin{figure}
    \centering
    \includegraphics[width=\linewidth]{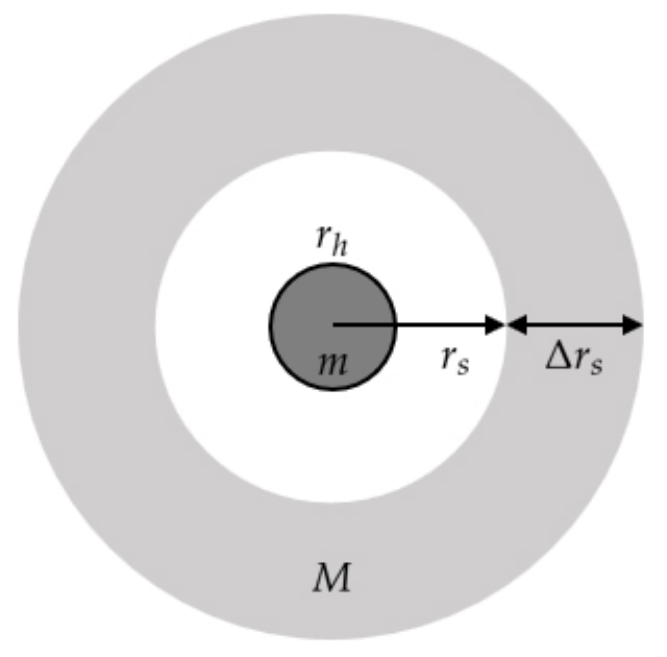}
    \caption{A representation of a black hole surrounded by a thin shell of dark matter where $r_{s} \neq r_{h}$.}
    \label{fig2}
\end{figure}

The general scenario is depicted in Fig. \ref{fig2} where $r_{s}\neq r_{h}$, which makes the piecewise function in Eq. \eqref{massfunc} very clear. Considering the first domain and if $r$ is between $r_{h}$ and $r_{s}$, the dark matter outside $r$ has no bearing to any black hole phenomena that we want to analyze. The situation is completely equivalent as if there's no dark matter surrounding the black hole. Suppose that $r$ describes the photonsphere radius $r_{ph}$, then we simply have the trivial expression $r_{ph}=3m$.

The third domain of Eq. \eqref{massfunc} seems an innocent-looking condition where the parameters $r_{s}$ and $\Delta r_{s}$ does not exist. The domain can be interpreted in two ways: (1) an isolated black hole surrounded by some dark matter shell, and (2) a black hole at the center of a galaxy (combined mass is $m$) where both are surrounded by dark matter halo. Taking the example of deriving the photonsphere radius, the former case gives $r_{ph}=3(m+M)$ as the null geodesic has no way to cross the outer radius of the dark matter shell. The study of the deflection angle is more applicable to the latter case, but the third domain's applicability rests on the assumption that the null geodesic never crosses the region of dark matter halo. For this to happen, the location of the source and the receiver must be tremendously remote from the lensing galaxy. It is also easy to see why the first and third domains should satisfy the Einstein equation.

For the second domain, the spacetime where the $r$-coordinate is located is now different due to the introduction of dark matter mass $M$ and its physical dimensions. Since the configuration is hypothetical, we don't know the specific field theory, which may lead to the stress-energy tensor that produces such expression of the metric if the Einstein equation is solved. It is also true for black hole spacetime fused with dark matter models whose density profiles came from observational data \cite{xu2018black,hou2018black, Jusufi2019}, hence they can be considered as dirty black holes. It is possible, however, to satisfy the Einstein equation. The key is to express the Einstein tensor in terms of appropriately chosen orthogonal bases along with the stress-energy tensor \cite{Azreg-Ainou2014}.

Assuming that the new spacetime expressed through the second domain of Eq. \eqref{massfunc} satisfies the Einstein equation, one can proceed with the analysis of black hole properties. It is found out that it gives a non-trivial expression for the photonsphere radius inside the dark matter shell, and is also true even with the approximation $\Delta r_{s}>>r_{s}$ \cite{Konoplya2019}. Nevertheless, it gave relevant insights to the effect of dark matter on the photonsphere radius, and as a consequence, to the deviations in the shadow radius. Note that being in the piecewise relation, the equation for the second domain cannot be applied for values of $r$ beyond its lower and upper bounds. An observer inside the dark matter shell has a different reality condition compared to being outside it, or as the first domain applies.

Looking at Eq. \eqref{e2} again, the function allows $r_{s}$ to have values smaller than $r = 2m$. For simplicity, one can set $r_{s}=2m$ and assume that the dark matter shell is static. By being static, it means that dark matter is not affected by the radial pull of the black hole. Nevertheless, it amplifies the change in the black hole's geometry, which then manifests to the dynamics of null and time-like particles.

The idea of an astrophysical environment that surrounds a black hole is common in several studies \cite{Konoplya2019a, Cardoso2017}. The reasons include (1) verify whether the deviation observed is due to the new physics happening near the horizon, or due to some effect of the astrophysical environment; and (2) directly examine and study the influence of astrophysical environment to the geometry of black hole. Only recently that the mass function in Eq. \eqref{massfunc} has been used to make some estimates of dark matter effects to the radius of the black hole shadow \cite{Konoplya2019}.

\section{Deflection angle in a dirty black hole using the Gauss-Bonnet theorem} \label{sec3}
\begin{figure}
    \centering
    \includegraphics[width=\linewidth]{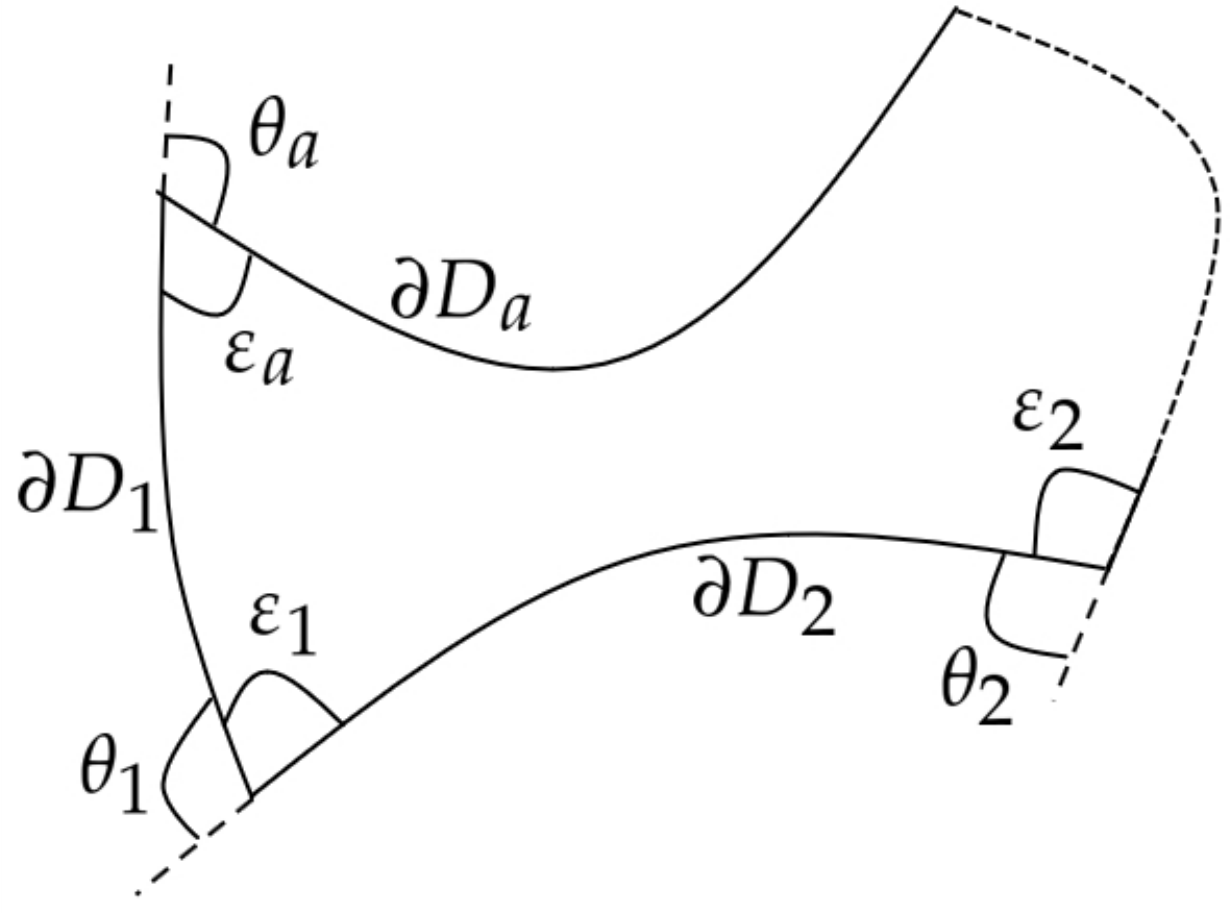}
    \caption{Schematic diagram of a curved surface for Gauss-Bonnet theorem. The interior angles are denoted by $\varepsilon$ while the exterior (or "jump") angles are $\theta$.}
    \label{gbtfig}
\end{figure}

Consider $D$ as a freely orientable two-dimensional curved surface described by the Gaussian curvature $K$, and $dS$ is its infinitesimal area element (See Fig. \ref{gbtfig}). It contains several boundaries that are differentiable curves, denoted by $\partial D_{a}$ ($a=1,2,...,N$), with geodesic curvature $\kappa_{g}$. Such boundary also contains the line element $\ell$ where its sign remains consistent with the surface orientation. Let $\theta_{a}$ represents the exterior or jump angles along the vertices of the surface. Then the Gauss-Bonnet theorem states that \cite{Ishihara2016,Ishihara2017,do2016differential,klingenberg2013course}
\begin{equation} \label{gbt}
    \iint_DKdS+\sum\limits_{a=1}^N \int_{\partial D_{a}} \kappa_{g} d\ell+ \sum\limits_{a=1}^N \theta_{a} = 2\pi.
\end{equation}

In the weak lensing regime, the Gauss-Bonnet theorem proves to be a powerful tool in calculating the weak deflection angle $\hat{\alpha}$ as long as the spacetime is asymptotically flat. If the spacetime is static, and spherically symmetric (SSS), $\kappa_{g}=0$ along the boundary curves and the second term in Eq. \eqref{gbt} vanishes. Assuming asymptotic flatness of the spacetime being considered, the weak deflection angle can be derived as \cite{Gibbons2008,Javed,Ishihara2016,Ishihara2017,Werner2012,Ono2017}
\begin{equation} \label{e5}
\hat{\alpha}=-\iint_{_{R}^{\infty }\square _{S}^{\infty
}}KdS 
\end{equation}
where $K$ is the Gaussian optical curvature that is integrated over the quadrilateral $_{R}^{\infty }\square _{S}^{\infty
}$, $dS$ is the surface area element. These quantities that are important in GBT, are defined as follows:
\begin{equation} \label{e6}
    K=\frac{R_{r\phi r\phi }}{\gamma },
\end{equation}
\begin{equation} \label{e7}
dS=\sqrt{\gamma}dr d\phi
\end{equation}
where $\gamma$ denotes the determinant of the spatial metric $\gamma_{ij}$ as $i$ and $j$ run from 1 to 3. The spatial metric for an SSS spacetime can be found by considering the null condition $ds^2=0$, and solving for $dt$ yields \cite{Ovgun2018}
\begin{equation} \label{e8}
    dt=\sqrt{\gamma_{ij}dx^i dx^j}.
\end{equation}
Rewriting Eq. \eqref{e3} as
\begin{equation} \label{e9}
   d{s}^2 = A(r)dt^2 +B(r)dr^2 + C(r) d\theta^2 + D(r,\theta) d\phi^2,
\end{equation}
the definition of $\gamma_{ij}$ is given by 
\begin{equation} \label{e10}
    \gamma_{ij}dx^i dx^j=\frac{1}{A(r)}\left(B(r)dr^2 + C(r) d\theta^2 + D(r,\theta) d\phi^2\right).
\end{equation}
Dealing only in equatorial orbits, the Gaussian curvature $K$, being related to the two-dimensional Riemann tensor, can be written in terms of affine connection as \cite{haroon2019shadow,Ovgun2019a}
\begin{equation} \label{e11}
    K=\frac{1}{\sqrt{\gamma}}\left[\frac{\partial}{\partial\phi}\left(\frac{\sqrt{\gamma}}{\gamma_{rr}}\Gamma_{rr}^{\phi}\right)-\frac{\partial}{\partial r}\left(\frac{\sqrt{\gamma}}{\gamma_{rr}}\Gamma_{r\phi}^{\phi}\right)\right]
\end{equation}
and using Eq. \eqref{e9}, $K$ can be written as \cite{Ono2017}
\begin{equation} \label{e12}
    K=-\sqrt{\frac{A(r)^{2}}{B(r)D(r)}}\frac{\partial}{\partial r}\left[\frac{1}{2}\sqrt{\frac{A(r)^{2}}{B(r)D(r)}}\frac{\partial}{\partial r}\left(\frac{D(r)}{A(r)}\right)\right].
\end{equation}
With the area surface element $dS$ in Eq. \eqref{e7}, we can write Eq. \eqref{e5} as
\begin{equation} \label{e13}
\hat{\alpha}=\int_{0}^{\pi}\int_{r_{o}}^{\infty}\mathcal{K}drd\phi,
\end{equation}
where the Gaussian curvature \textit{term} \cite{DeLeon2019} reads
\begin{equation} \label{e14}
    \mathcal{K}=-\frac{\partial}{\partial r}\left[\frac{1}{2}\sqrt{\frac{A(r)^{2}}{B(r)D(r)}}\frac{\partial}{\partial r}\left(\frac{D(r)}{A(r)}\right)\right].
\end{equation}
Also, $r_{o}$ denotes the radial distance of the photon's closest approach to the lensing black hole, and can be obtained through the orbit equation. For an SSS metric such as the Schwarzschild metric, the photon's orbit equation in the equatorial plane reads
\begin{equation} \label{n15}
    \left(\frac{dr}{d\phi}\right)^2=\frac{D(r)(D(r)-A(r)b^2)}{A(r)B(r)b^2}
\end{equation}
where $b$ is the impact parameter. For convenience and as usually done in celestial mechanics, we can set $u=\frac{1}{r}$. Thus, the above can be expressed as
\begin{equation} \label{n16}
    \left(\frac{du}{d\phi}\right)^2 \equiv F(u)=\frac{u^4D(u)(D(u)-A(u)b^2)}{A(u)B(u)b^2}.
\end{equation}
The closest approach $u_{o}$ can be found by iteratively solving $u$ in Eq. \eqref{n16} and imposing the boundary condition $\frac{du}{d\phi}\big|_{\phi=\frac{\pi}{2}}=0$. For example, for the Schwarzschild metric, $u$ is given by
\begin{equation} \label{n17}
   u=\frac{\sin \phi}{b}+\frac{m}{b^2}(1+\cos^2 \phi).
\end{equation}
With $u_{o}$, we can now recast Eq. \eqref{e13} as
\begin{equation} \label{n18}
\hat{\alpha}=\int_{0}^{\pi}\int_{0}^{u_{o}}-\frac{\mathcal{K}}{u^2}dud\phi.
\end{equation}

\subsection{Case 1: $\mathcal{M}(r)=m$}
\begin{figure}
    \centering
    \includegraphics[width=\linewidth]{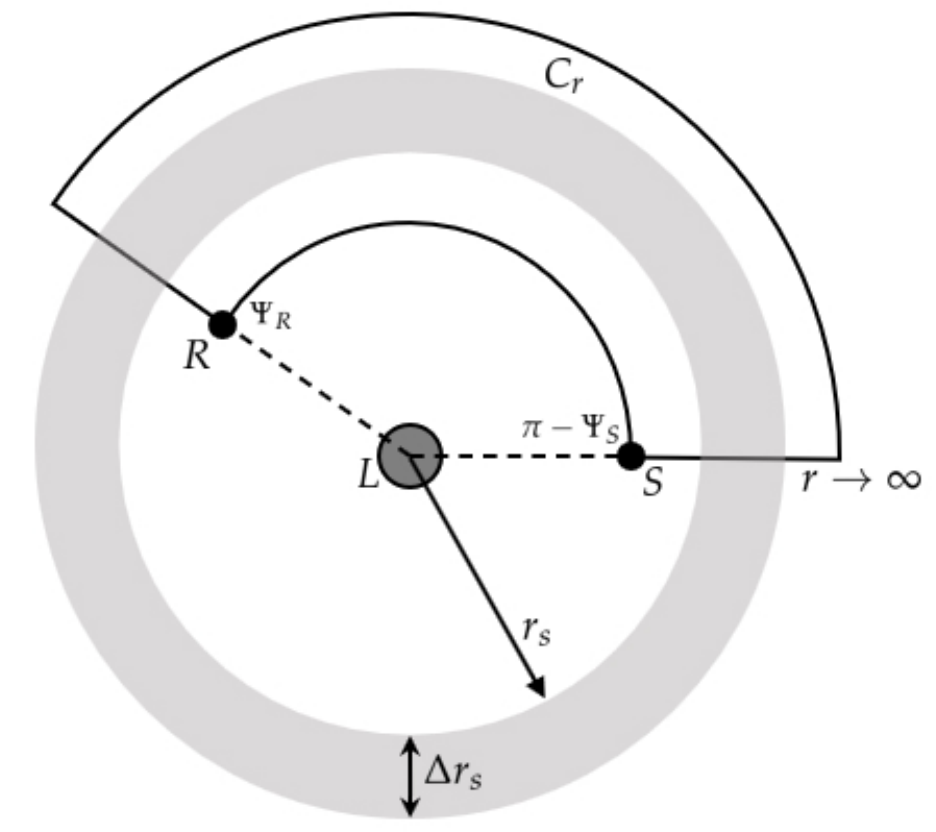}
    \caption{Quadrilateral embedded in curved spacetime while enclosing dark matter shell.}
    \label{fig3}
\end{figure}

Let $u_{S}$ and $u_{R}$ be the reciprocal of the source's and receiver's radial distance from the lensing object. The use of Eq. \eqref{e13} assumes that these radial distances are so far away that the spacetime becomes approximately Minkowskian. For the first domain in Eq. \eqref{massfunc} to hold, the inner radius $r_{s}$ of the dark matter shell should always larger than the source's and receiver's radial distance (See Fig. \ref{fig3}). With the metric function taking the form
\begin{equation} \label{e15}
    f(r)=1-\frac{2m}{r},
\end{equation} 
the Gaussian curvature and the area element in the equatorial plane reads
\begin{align} \label{n20}
    K&=-\frac{2m}{r^3}+\frac{3m^2}{r^4} \nonumber \\
    dS&=r+3m+\mathcal{O}(m^2)
\end{align}
respectively. Using Eq. \eqref{n18}, we find
\begin{align} \label{e16}
   \hat{\alpha}&=\int_{0}^{\pi}\int_{0}^{\frac{\sin(\phi)}{b}+\frac{m}{b^2}(1+\cos^2 \phi)}-\frac{\mathcal{K}}{u^2}du d\phi \nonumber \\
   &=\int_{0}^{\pi}\int_{0}^{\frac{\sin(\phi)}{b}+\frac{m}{b^2}(1+\cos^2 \phi)} 2m+3m^2u +\mathcal{O}\left(m^3\right) du d\phi \nonumber \\
   &=\frac{4m}{b}+\frac{15\pi m^2}{4b^2}+\mathcal{O}\left(m^3\right)
\end{align}
which is the weak deflection angle of a Schwarzschild black hole up to the second-order in $m$.

\subsection{Case 2: $\mathcal{M}(r)=m+M$}
\begin{figure}
    \centering
    \includegraphics[width=\linewidth]{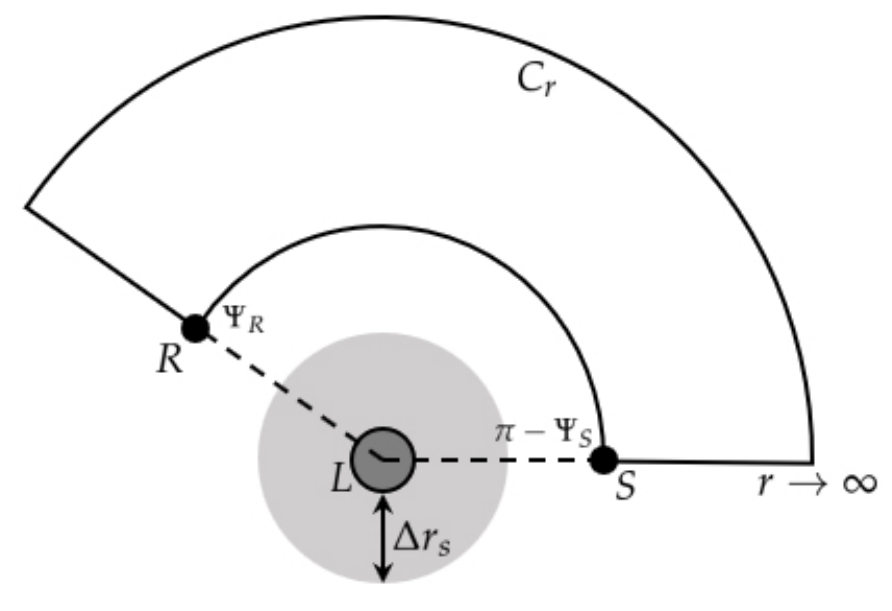}
    \caption{Geometrical configuration of the quadrilateral when S and R are beyond the outermost radius of dark matter shell.}
    \label{fig4}
\end{figure}

Following the discussion in Sect. \ref{sec2} about the possible interpretation of this case, Fig. \ref{fig4} shows the use of the third domain of Eq. \eqref{massfunc}. As we deal only with a non-rotating black hole, the metric function is trivial:
\begin{equation} \label{e17}
    f(r)=1-\frac{2(m+M)}{r}.
\end{equation}
Using Eq. \eqref{e13}, we find
\begin{align} \label{e18}
   \hat{\alpha}&=\int_{0}^{\pi}\int_{0}^{\frac{\sin(\phi)}{b}+\frac{m+M}{b^2}(1+\cos^2 \phi)}-\frac{\mathcal{K}}{u^2}du d\phi \nonumber \\
   &=\int_{0}^{\pi}\int_{0}^{\frac{\sin(\phi)}{b}+\frac{m+M}{b^2}(1+\cos^2 \phi)} 2(m+M) \nonumber \\
   &+3(m+M)^2u +\mathcal{O}\left[(m+M)^3\right] du d\phi \nonumber \\
   &=\frac{4(m+M)}{b}+\frac{15\pi (m+M)^2}{4b^2}+\mathcal{O}\left[(m+M)^3\right]
\end{align}

The result in Eq. \eqref{e18} shows clearly how the dark matter mass acts as an additional effective mass to the black hole - it increases the value of the weak deflection angle. It clearly shows how $M$ adds up to the spacetime distortion that affects the path of the null geodesic. It also means that the photon's path never enters and leaves the dark matter shell as it travels from the source to the receiver. Suppose that $M=0$ and we increase $m$, the null geodesic would only shift to its new orbital equilibrium while being outside the vicinity of the event horizon.

\subsection{Case 3: $\mathcal{M}(r)=m+MG(r)$}
In this scenario, the null geodesic should remain inside the dark matter shell for the second domain in Eq. \eqref{massfunc} to hold. Such restriction is to determine how the weak deflection angle should behave, knowing that the dark matter beneath or above the null geodesic can affect it. This scenario is the same as the restriction imposed in Ref. \cite{Konoplya2019} for the photonsphere radius.

\begin{figure}
    \centering
    \includegraphics[width=\linewidth]{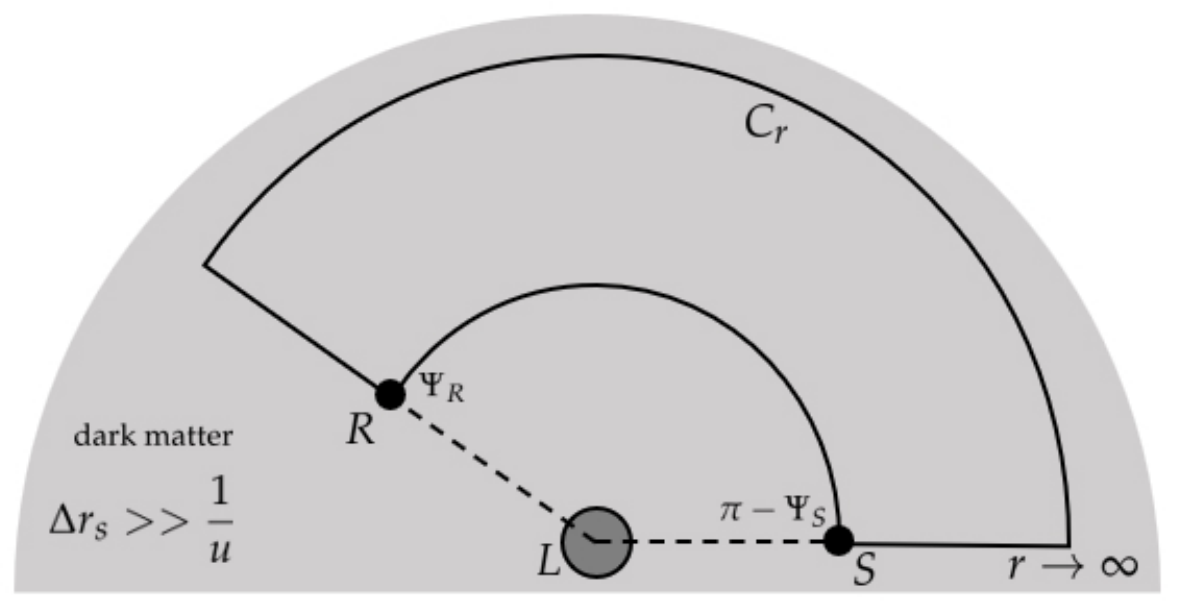}
    \caption{Far approximation of S and R (i.e. $u_{S}<<1$ and $u_{S}<<1$) always implies the approximation that $\Delta r_{s}>>\frac{1}{u_{S}}$ (or $u_{R}$) for the 2nd domain in Eq. \eqref{massfunc} to hold.}
    \label{fig5}
\end{figure}

Suppose that $\Delta r_{s}$ is fixed where its thickness is not yet comparable to dark matter halos and accommodates both the source and the receiver inside the shell. In calculating the weak deflection angle using Eq. \eqref{n18}, the asymptotic condition must apply, but the problem is that the source and the receiver locations exceed that of the shell's outer radius. Hence, given Eq. \eqref{massfunc}, such a situation invalidates the use of Eq. \eqref{e2}. Therefore, there is a necessity for the approximation of $\Delta r_{s}$ that it must be very large. Such approximation directs the analysis to the scenario involving the source and the receiver inside a very large dark matter halo. This means there is an assumption that $\Delta r_{s}>>\frac{1}{u_{S}}$ (and $u_{R}$). See Fig. \ref{fig5}.

The orbit equation in this case is
\begin{align} \label{n24}
   F(u)&=\frac{1}{b^{2}}+u^{2}(2mu-1) \nonumber \\
   &+\frac{2M}{\Delta r_{s}^{2}}(r_{s}u-1)^{2}\left(\frac{2r_{s}u}{\Delta r_{s}}-\frac{2}{\Delta r_{s}}+3u\right).
\end{align}
Note that there is no coupling between the black hole mass $m$ and dark matter mass $M$ in Eq. \eqref{n24}, unlike the coupling between the spin parameter $a$ and $m$ in Kerr spacetime. The iterative solution of Eq. \eqref{n24} to obtain $u_{o}$ is then
\begin{equation} \label{n25}
    u=\frac{\sin \phi}{b}+\frac{m}{b^2}(1+\cos^2 \phi)+\frac{3Mr_{s}^2}{b^2\Delta r_{s}^2}.
\end{equation}
Hence, the integrand of Eq. \eqref{n18}, up to the third-order of $1/\Delta r_{s}$, is expressed by
\begin{align} \label{e19}
   \hat{\alpha}&=\int_{0}^{\pi}\int_{0}^{u_{o}}-\frac{\mathcal{K}}{u^2}du d\phi=\int_{0}^{\pi}\int_{0}^{u_{o}}\bigg[2m+\frac{6Mr_{s}^{2}}{\Delta r_{s}^{2}} \nonumber \\
   &-\frac{6mMr_{s}(2-3r_{s}u)}{\Delta r_{s}^{2}}-\frac{4M}{\Delta r_{s}^{3}}\left(\frac{1}{u^{3}}-r_{s}^{3}\right) \nonumber \\
   &-\frac{12mMr_{s}^{2}(1-r_{s}u)}{\Delta r_{s}^{3}}+\mathcal{O}\left(m^2,\frac{1}{\Delta r_{s}^4}\right)\bigg] du d\phi.
\end{align}
The result shows divergence in the 4th term if one attempts to evaluate the integral. It means that in the third-order of $1/\Delta r_{s}$, if one insists to be more precise in their calculations, the spacetime is not asymptotically flat and GBT cannot be used. The same case also occur in other non-asymptotic spacetime such as those that involved the cosmological constant such as the Kottler spacetime \cite{doi:10.1002/andp.19183611402}. However, the cosmological constant has no adjustable parameters where an approximation can be made possible.

The condition $\Delta r_{s}>>\frac{1}{u}$ gives the approximate expression if $\Delta r_{s}$ is very large, at least comparable to the known size of dark matter halos. Therefore, the manifestation of dark matter effects to the weak deflection angle can be seen to begin with the second-order of $1/\Delta r_{s}$. Fortunately, there is asymptotic flatness and proceeding with the GBT calculation, we find
\begin{align} \label{e20}
   \hat{\alpha}&=\int_{0}^{\pi}\int_{0}^{u_{o}}-\frac{\mathcal{K}}{u^2}du d\phi=\int_{0}^{\pi}\int_{0}^{u_{o}}\bigg[2m+\frac{6Mr_{s}^{2}}{\Delta r_{s}^{2}} \nonumber \\
   &-\frac{6mMr_{s}(2-3r_{s}u)}{\Delta r_{s}^{2}}\bigg] du d\phi=\frac{4m}{b}+\frac{12Mr_{s}^{2}}{b\Delta r_{s}^{2}}\nonumber \\
   &-\frac{24mMr_{s}}{b\Delta r_{s}^{2}}+\frac{15\pi m^2}{4b^2}+\frac{39\pi mMr_{s}^{2}}{2b^{2}\Delta r_{s}^2}.
\end{align}
showing higher-order terms. It is now a matter of question whether the third term in Eq. \eqref{e20} can also be neglected due to the nature of dark matter model used in this study. To find out, we will use Ishihara (\textit{et al.}) method \cite{Ishihara2016} to purposely show that we can compute the weak deflection angle up to the third-order in $1/\Delta r_{s}$, and compare the result to Eq. \eqref{e20} at least in second-order of $1/\Delta r_{s}$. The formalism developed is very useful in dealing with Kottler spacetime and Schwarzschild-like solutions in the Weyl conformal gravity \cite{mannheim1989exact} because it assumes that $u_{S}$ and $u_{R}$ are at finite distance from the black hole. The computation using this method will be the subject of the next section.

\section{Deflection angle using finite distance} \label{sec4}
In this section, we use the method in Ref. \cite{Ishihara2016} to calculate the weak deflection angle of a black hole under the influence of dark matter, as demonstrated in Fig. \ref{fig5}. From the GBT, the generalized correspondence between the deflection angle and the surface integral of the Gaussian curvature reads
$$\hat{\alpha}=\phi_{RS}+\Psi_{R}-\Psi_{S}$$
\begin{equation} \label{e21}
    =\int_{u_{R}}^{u_{o}}\frac{1}{\sqrt{F(u)}}du+\int_{u_{S}}^{u_{o}}\frac{1}{\sqrt{F(u)}}du+\Psi_{R}-\Psi_{S}
\end{equation}
where $F(u)$ and $u_{o}$ are given by Eq. \eqref{n24} and Eq. \eqref{n25} respectively. The angles $\Psi_{S}$ and $\Psi_{R}$ in Eq. \eqref{e21} are determined through the inner product of the unit basis vector in the spacetime considered, and the unit vector with respect to the lensing object. The unit basis vector $e^i$, along the equatorial plane, is given by
\begin{equation} \label{e24}
    e^i=\left(\frac{dr}{dt},0,\frac{d\phi}{dt}\right)=\frac{d\phi}{dt}\left(\frac{dr}{d\phi},0,1\right)
\end{equation}
while the unit radial vector, which is along the radial direction from the lens is
\begin{equation} \label{e25}
    R^i=\left(\frac{1}{\sqrt{\gamma_{rr}}},0,0\right).
\end{equation}
Hence, the inner product suggests the definition
$$\cos \Psi\equiv\gamma_{ij}e^iR^j$$
\begin{equation} \label{e26}
    \cos \Psi=\sqrt{\gamma_{rr}}\frac{A(r)b}{D(r)}\frac{dr}{d\phi}.
\end{equation}
Using $F(u)$ in Eq. \eqref{n16}, it is easy to see that
\begin{equation} \label{e27}
    \sin\Psi=\sqrt{\frac{A(r)}{D(r)}}b
\end{equation}
where it clear that it is favorable to use $\sin\Psi$ than using $\cos\Psi$.

Taylor expansion of $\Delta r_{s}$ as it is very large resulted to an angle $\Psi$ calculated as
$$\Psi=\arcsin(bu\sqrt{1-2mu})$$
$$-\frac{3bM}{\Delta r_{s}^2}\frac{(r_{s} u-1)^2}{\sqrt{1-2 m u} \sqrt{b^2 u^2 (2 m u-1)+1}}$$
\begin{equation} \label{e28}
    -\frac{2bM}{\Delta r_{s}^3}\frac{(r_{s} u-1)^3}{u \sqrt{1-2 m u} \sqrt{b^2 u^2 (2 m u-1)+1}}+\mathcal{O}\left(\frac{1}{\Delta r_{s}^4}\right).
\end{equation}
Continuing, we find that
\begin{widetext}
$$\Psi_{R}-\Psi_{S}=(\Psi_{R}^{\text{Schw}}-\Psi_{S}^{\text{Schw}})-\frac{3bM}{\Delta r_{s}^2}\left[\frac{(r_{s}u_{R}-1)^{2}}{\sqrt{1-b^{2}u_{R}^{2}}}+ \frac{(r_{s}u_{S}-1)^{2}}{\sqrt{1-b^{2}u_{S}^{2}}}\right]$$
$$-\frac{2bM}{\Delta r_{s}^{3}}\left[\frac{(r_{s}u_{S}-1)^{3}}{u_{R}\sqrt{1-b^{2}u_{R}^{2}}}+\frac{(r_{s}u_{S}-1)^{3}}{u_{S}\sqrt{1-b^{2}u_{S}^{2}}}\right]$$
$$+\frac{3bmM}{\Delta r_{s}^{2}}\left[\frac{u_{R}\left(2b^{2}u_{R}^{2}-1\right)(r_{s}u_{R}-1)^{2}}{\left(1-b^{2}u_{R}^{2}\right)^{3/2}}+\frac{u_{S}\left(2b^{2}u_{S}^{2}-1\right)(r_{s}u_{S}-1)^{2}}{\left(1-b^{2}u_{S}^{2}\right)^{3/2}}\right]$$
\begin{equation} \label{e29}
    +\frac{2bmM}{\Delta r_{s}^{3}}\left[\frac{\left(2b^{2}u_{R}^{2}-1\right)(r_{s}u_{R}-1)^{3}}{\left(1-b^{2}u_{R}^{2}\right)^{3/2}}+\frac{\left(2b^{2}u_{S}^{2}-1\right)(r_{s}u_{S}-1)^{3}}{\left(1-b^{2}u_{S}^{2}\right)^{3/2}}\right]+\mathcal{O}\left(\frac{1}{\Delta r_{s}^4}\right)
\end{equation}
\end{widetext}
where
$$(\Psi_{R}^{\text{Schw}}-\Psi_{S}^{\text{Schw}})=[\arcsin(u_{R}b)+\arcsin(u_{S}b)-\pi]$$
\begin{equation} \label{e30}
    -bm\left[\frac{u_{R}^{2}}{\sqrt{1-b^{2}u_{R}^{2}}}+ \frac{u_{S}^{2}}{\sqrt{1-b^{2}u_{S}^{2}}}\right].
\end{equation}
Note the above expression contains a divergent term in the third-order of $1/\Delta r_{s}$. Hence, the spacetime caused by the combination of black hole and dark matter with very large $\Delta r_{s}$ is non-asymptotic and the limit $u_{S}\rightarrow0$ and $u_{R}\rightarrow0$ is not allowed.

We now proceed to calculate the $\phi_{RS}$ part by evaluating the integrals in Eq. \eqref{e21}. The function $F(u)$ in Eq. \eqref{n24} results to an integrand of the form
\begin{widetext}
$$\frac{1}{\sqrt{F(u)}}=\frac{b}{\sqrt{1-b^{2}u^{2}}}-\frac{b^{3}u^{3}m}{\left(1-b^{2}u^{2}\right)^{3/2}}$$
$$-\frac{3Mb^3}{\Delta r_{s}^2}\frac{u(r_{s}u-1)^{2}}{\left(1-b^{2}u^{2}\right)^{3/2}}+\frac{9b^{5}mMu^{4}}{\Delta r_{s}^2}\frac{(r_{s}u-1)^{2}}{\left(1-b^{2}u^{2}\right)^{5/2}}$$
\begin{equation} \label{e31}
    -\frac{2Mb^3}{\Delta r_{s}^3}\frac{(r_{s}u-1)^{3}}{\left(1-b^{2}u^{2}\right)^{3/2}}+\frac{6b^{5}mMu^{3}}{\Delta r_{s}^3}\frac{(r_{s}u-1)^{3}}{\left(1-b^{2}u^{2}\right)^{5/2}}+\mathcal{O}\left(\frac{1}{\Delta r_{s}^4}\right).
\end{equation}
\end{widetext}
The evaluation of the integral gives a cumbersome expression and reveals no divergence especially in the third-order of $1/\Delta r_{s}$ and hence, can be safely omitted for brevity. Due to Eq. \eqref{e21}, the expression for the source and receiver are essentially the same, and that is
\begin{widetext}
$$\int\frac{1}{\sqrt{F(u)}}du=\arcsin bu+\frac{m}{b}\frac{\left(b^{2}u^{2}-2\right)}{\sqrt{1-b^{2}u^{2}}}$$
$$+\frac{9M}{\Delta r_{s}^2}\left[\left(m-\frac{2r_{s}}{3}\right)+\frac{5mr_{s}^{2}}{2b^{2}}\right]\arcsin(bu)-\frac{3M}{b\Delta r_{s}^2}\frac{\left[b^{2}\left(-r_{s}^{2}u^{2}-2r_{s}u+1\right)+2r_{s}^{2}\right]}{\sqrt{1-b^{2}u^{2}}}$$
\begin{equation} \label{e32}
    -\frac{3mMr_{s}}{2b\Delta r_{s}^2}\frac{\left(48r_{s}b^{2}u^{2}+6b^2u+15r_{s}^2u-32r_{s}\right)}{\left(1-b^{2}u^{2}\right)^{3/2}}+\mathcal{O}\left(\frac{1}{\Delta r_{s}^3}\right)+C
\end{equation}
where C is a constant. The expression for $\phi_{RS}$ includes the sum of two evaluated integrals:
$$\phi_{RS}=\phi_{RS}^{\text{Schw}}-\frac{9M}{\Delta r_{s}^{2}}\left[\left(m-\frac{2r_{s}}{3}\right)+\frac{5mr_{s}^{2}}{2b^{2}}\right]\left[\arcsin(bu_{R})+\arcsin(bu_{S})\right]$$
$$+\frac{3M}{b\Delta r_{s}^{2}}\left\{ \frac{\left[b^{2}\left(-r_{s}^{2}u_{R}^{2}-2r_{s}u_{R}+1\right)+2r_{s}^{2}\right]}{\sqrt{1-b^{2}u_{R}^{2}}}+\frac{\left[b^{2}\left(-r_{s}^{2}u_{S}^{2}-2r_{s}u_{S}+1\right)+2r_{s}^{2}\right]}{\sqrt{1-b^{2}u_{S}^{2}}}\right\}$$
\begin{align} \label{e33}
    &+\frac{3mMr_{s}}{2b\Delta r_{s}^{2}}\bigl[\frac{\left(48r_{s}b^{2}u_{R}^{2}+6b^2u_{R}+15r_{s}^2u_{R}-32r_{s}\right)}{\left(1-b^{2}u_{R}^{2}\right)^{3/2}} \nonumber \\
    &+\frac{\left(48r_{s}b^{2}u_{S}^{2}+6b^2u_{S}+15r_{s}^2u_{S}-32r_{s}\right)}{\left(1-b^{2}u_{S}^{2}\right)^{3/2}}\bigr]+\mathcal{O}\left(\frac{1}{\Delta r_{s}^3}\right)
\end{align}
where we introduced
\begin{equation} \label{e34}
    \phi_{RS}^{\text{Schw}}=\pi-\arcsin(u_{R}b)-\arcsin(u_{S}b)-\frac{m}{b}\left[\frac{\left(b^{2}u_{R}^{2}-2\right)}{\sqrt{1-b^{2}u_{R}^{2}}}+ \frac{\left(b^{2}u_{S}^{2}-2\right)}{\sqrt{1-b^{2}u_{S}^{2}}}\right].
\end{equation}
We can finally calculate the deflection angle $\hat{\alpha}$ using Eq. \eqref{e21}. Combining Eq. \eqref{e29} and Eq. \eqref{e33}, we find
$$\hat{\alpha}\approx\frac{2m}{b}\left[\sqrt{1-b^{2}u_{R}^{2}}+\sqrt{1-b^{2}u_{S}^{2}}\right]$$
$$-\frac{9M}{\Delta r_{s}^{2}}\left[\left(m-\frac{2r_{s}}{3}\right)+\frac{5mr_{s}^{2}}{2b^{2}}\right]\left[\arcsin(bu_{R})+\arcsin(bu_{S})\right]$$
$$+\frac{6Mr_{s}^{2}}{b\Delta r_{s}^{2}}\left[\sqrt{1-b^{2}u_{R}^{2}}+\sqrt{1-b^{2}u_{S}^{2}}\right]-\frac{2bM}{\Delta r_{s}^3}\left[\frac{(r_{s}u_{R}-1)^{3}}{u_{R}\sqrt{1-b^{2}u_{R}^{2}}}+\frac{(r_{s}u_{S}-1)^{3}}{u_{S}\sqrt{1-b^{2}u_{S}^{2}}}\right]-$$
$$\frac{3mM}{2\Delta r_{s}^{2}}\left\{ \frac{\left[4b^2u_{R}+r_{s}(52b^2u_{R}^2+15r_{s}u_{R}-32)\right]}{b\left(1-b^{2}u_{R}^{2}\right)^{3/2}}+\frac{\left[4b^2u_{S}+r_{s}(52b^2u_{S}^2+15r_{s}u_{S}-32)\right]}{b\left(1-b^{2}u_{S}^{2}\right)^{3/2}}\right\}$$
\begin{equation} \label{e35}
    +\frac{2bmM}{\Delta r_{s}^3}\left[\frac{\left(2b^{2}u_{R}^{2}-1\right)(r_{s}u_{R}-1)^{3}}{\left(1-b^{2}u_{R}^{2}\right)^{3/2}}+\frac{\left(2b^{2}u_{S}^{2}-1\right)(r_{s}u_{S}-1)^{3}}{\left(1-b^{2}u_{S}^{2}\right)^{3/2}}\right]+\mathcal{O}\left(\frac{1}{\Delta r_{s}^4}\right).
\end{equation}
\end{widetext}

\begin{figure}
    \centering
    \includegraphics[width=\linewidth]{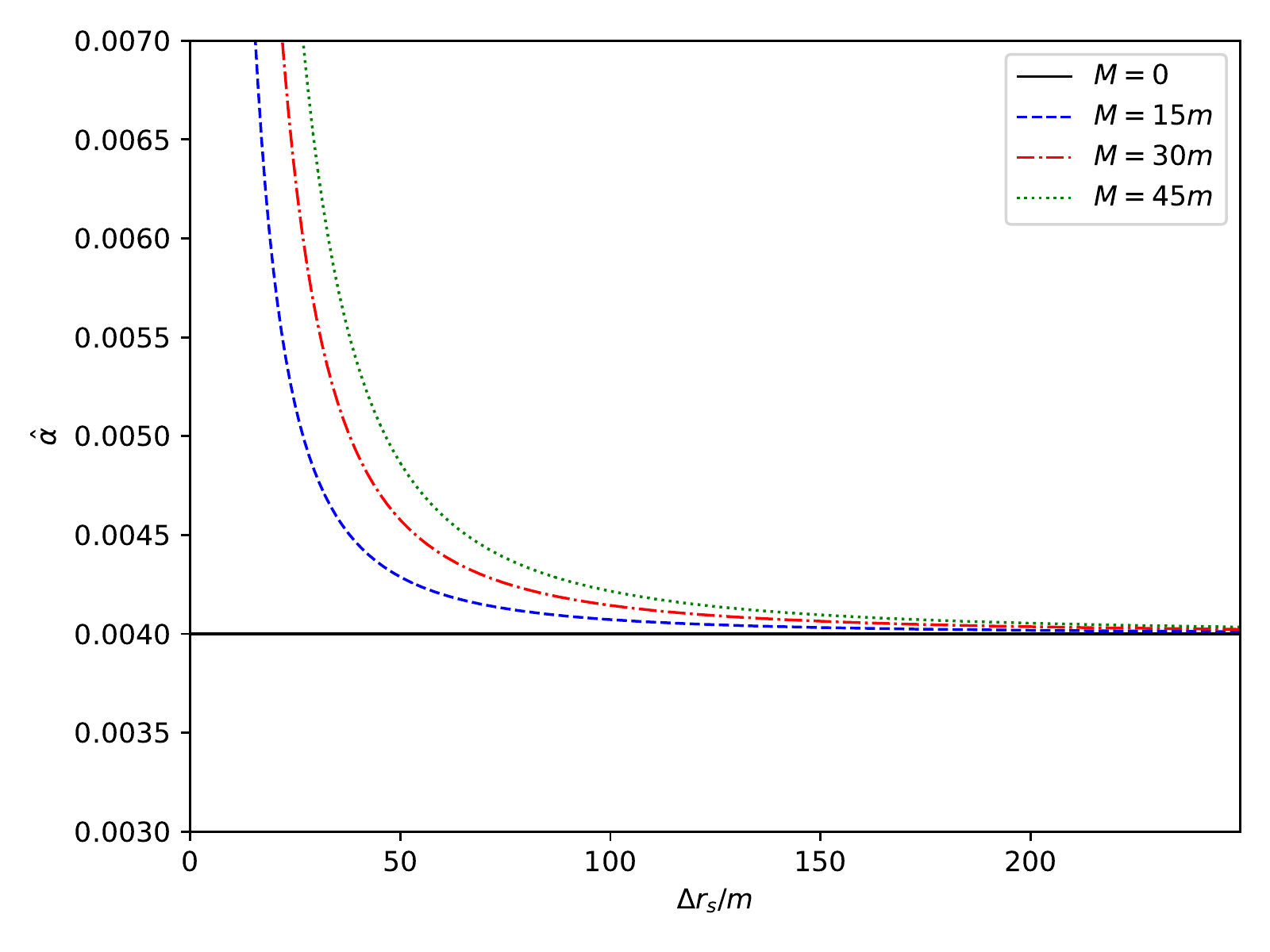}
    \caption{Weak deflection angle as $\Delta r_{s}$ varies. Here, $r_{s}=2m$ which coincides with the event horizon, and $b=1000m$.}
    \label{fig6}
\end{figure}

The next procedure is to impose the limit $u_{S}\rightarrow0$ and $u_{S}\rightarrow0$ in Eq. \eqref{e35} which is not problematic in asymptotic spacetimes. However, this limit is not applicable or allowed in Eq. \eqref{e35} since $\hat{\alpha}$ will apparently diverge. Hence, the reciprocal of $u_{S}$ and $u_{R}$ can be interpreted as finite distances but with very small values. For the far source and receiver, it is then safe to impose $u_{S}<<1$ and $u_{R}<<1$ \cite{Ishihara2016,haroon2019shadow}. Eq. \eqref{e35} then becomes
$$\hat{\alpha}\approx\frac{4m}{b}+\frac{12Mr_{s}^2}{b\Delta r_{s}^2}-\frac{96mMr_{s}}{b\Delta r_{s}^2}$$
\begin{equation} \label{e36}
    +\frac{2Mb}{\Delta r_{s}^{3}}\left[\left(\frac{1}{u_{R}}+\frac{1}{u_{S}}\right)+2m\right]+\mathcal{O}\left(\frac{1}{\Delta r_{s}^4}\right).
\end{equation}
Comparing this result to Eq. \eqref{e20}, we see that the first two terms are identical, except the third term, and the existence of terms in the third-order of $1/\Delta r_{s}$ which cannot be derived from the GBT given by Eq. \eqref{e13}. Dark matter mass, being an additional effective mass, makes the $mM$ term to be interpreted as a higher-order term in mass and hence, can safely be neglected. The significant contribution to the weak deflection angle only arises from the first two terms. Moreover, the source and the receiver are inside the dark matter halo, hence $\Delta r_{s}>>1/u_{S}$ and $\Delta r_{s}>>1/u_{R}$ is possible. Hence, we are left with
\begin{equation} \label{e37}
    \hat{\alpha}\approx\frac{4m}{b}+\frac{12Mr_{s}^2}{b\Delta r_{s}^2}
\end{equation}
and we see that dark matter has an effect to increase the value of the weak deflection angle. See Fig. \eqref{fig6}. Moreover, the abnormality of such increase in the weak deflection angle due to the very small $\Delta r_{s}$ might constrain the value of dark matter mass $M$. The weak deflection angle is asymptotic to the Schwarzschild case as $\Delta r_{s}$ increases, as expected. Note that if the mass $M$ is negative, which might represent an exotic matter with negative kinetic term, the weak deflection angle decreases.

We recall the estimate in Ref. \cite{Konoplya2019} that for notable dark matter effects to occur in the shadow radius, the effective radius of the dark matter distribution must be in the order of $\Delta r_{s}\sim\sqrt{3mM}$. This result is found by the same analysis using Fig. \ref{fig5}. For a given value of $M$, it revealed a very small value of $\Delta r_{s}$, which implies that dark matter must be concentrated near the black hole to change the shadow radius considerably. Further analysis also revealed that when the dark matter density is very low ($\Delta r_{s}$ is very large), the effect of dark matter outside the photon's orbit can be safely neglected. Hence, the dark matter mass beneath the photon's orbit is the main contributor to any deviations. Such a previous conclusion synchronizes with the results in Eq. \eqref{e20} and Eq. \eqref{e36}. Since dark matter acts as an additional effective mass to the black hole, we can interpret the $mM$ term as a higher-order term in mass, and their coupling can be ignored.

Surprisingly, Eq. \eqref{e37} reveals that dark matter effects will occur in the weak deflection angle (as deviation) when $\Delta r_{s}=2\sqrt{3mM}$. Since there is an increase in the thickness requirement, dark matter detection via deviations in the weak deflection angle of a black hole is better compared when a deviation in shadow radius is used. Unfortunately, however, such an increase is still irrelevant, at least for the technological capabilities we have today. Consider an estimate for dark matter mass in our galaxy which is $M\approx1.0\times10^{12}M_{\odot}$ \cite{Battaglia2005} while the mass of the central black hole is around $m\approx4.3\times10^{6}M_{\odot}$. It gives a required dark matter thickness of $\Delta r_{s}\approx6.13\times10^{12}$ m $\approx2\times10^{-4}$ pc to see any changes in the weak deflection angle. Such value is lower in many orders of magnitude even if we compare it to the core radius of the dark matter halo present in our galaxy ($r_{o}\approx15.7-17.46$ kpc) \cite{DeOliveira2015}.

\section{Conclusion} \label{sec5}
In this paper, we present an analytic formula for the weak deflection angle using a simple dark matter model that only incorporates its basic features, such as mass and physical parameters. It is shown that the GBT, with its form given by Eq. \eqref{n18}, can be used only to the second-order of $1/\Delta r_{s}$, but is ill-behaved in the third-order of $1/\Delta r_{s}$. If precision in the calculation matters, the resulting apparent divergence is found and justified using Ishihara (\textit{et al.}) method. The expression containing the second-order in $1/\Delta r_{s}$ represents the approximate condition where the initial manifestation of dark matter effect occurs as a deviation in the weak deflection angle. We found that it is twice that of shadow radius deviations. Interestingly, if one seeks to detect dark matter using the central black hole in one's galaxy, using the weak deflection angle is better than observing deviations in the shadow radius. Although being better, the deviation is still very small to be detected by current technology.

Extensions of the current study to non-spherical, or non-static dark matter distribution, surrounding a more complicated black hole metrics, are left for future work.

\bibliography{references}

\end{document}